\documentclass[aps,preprint,superscriptaddress,prb]{revtex4}
\usepackage{epsfig}
\usepackage{amsmath}
\usepackage{graphicx}
\usepackage{color}
\usepackage{color}
\usepackage{ulem}

\newcommand{\ud}{{\mathrm d}}

\begin{document}

\title{Kinetics of intermediate-mediated self-assembly in nano-sized
materials: a generic model}
\author{James F. Lutsko}
\email{E-mail: jlutsko@ulb.ac.be}
\author{Vasileios Basios}
\author{Gr\'{e}goire Nicolis}
\affiliation{Center for Nonlinear Phenomena and Complex Systems CP 231, Universit\'{e}
Libre de Bruxelles, Blvd. du Triomphe, 1050 Brussels, Belgium}
\author{Tom P. Caremans}
\author{Alexander Aerts}
\author{Johan A. Martens}
\author{Christine E. A. Kirschhock}
\author{Titus S. van Erp}
\email{E-mail: Titus.VanErp@biw.kuleuven.be}
\affiliation{Center for Surface Chemistry and Catalysis, K. U. Leuven, Kasteelpark
Arenberg 23, 3001 Leuven, Belgium}
\date{\today }

\begin{abstract}
We propose in this paper a generic model of a non-standard aggregation
mechanism for self-assembly processes of a class of materials involving the
mediation of intermediates consisting of a polydisperse population of
nano-sized particles. The model accounts for a long induction period in the
process. The proposed mechanism also gives insight on future experiments
aiming at a more comprehensive picture of the role of self-organization in
self-assembly processes.
\end{abstract}

\maketitle

\renewcommand{\thefootnote}{\fnsymbol{footnote}} \renewcommand{%
\theequation}{\arabic{section}.\arabic{equation}}

\section{Introduction}

\label{secintro}

Nanophase materials are of key importance in a variety of fields from molecular and cellular biology to technological innovation. They
also pose major theoretical challenges in view of their multiple scale
dynamics, whereby microscopic level processes determine macroscopic
properties through the growth of initial fluctuations favoring in some way
the specific material that will eventually emerge from synthesis. This
switches on, in turn, a variety of self-organization phenomena associated
with the coexistence of competing pathways.

The processes we investigate in this work involve two steps. First, an
initial species, N, is able to agglomerate into an intermediate form, X. We
will primarily envisage the increase/decrease in the amount of X as being
due to the attachment/detachment of units of N although more complex
dynamics might be relevant in some cases. We expect that N and X will reach
an equilibrium in a relatively short time. The second step of the process is
the self-assembly of the final product, S, from X. By ``self-assembly'' we
imagine that the units of X must not only join together to form S, but must
undergo some sort of internal transition (such as a restructuring) which is
quite slow. One of the main points of this work is to show that simply
assuming a long characteristic time (i.e. a small rate constant) for this
internal transition is not sufficient to give a long induction time.
However, by including cooperativity - whereby the nascent S material
templates or catalyzes the internal transition - arbitrarily long induction
times can be achieved. A mathematical model of the contribution of
cooperativity will be developed based on general physical considerations of
how the process must proceed.

The need to synthesize intermediate substances prior to the material of
interest is also a well known phenomenon in chemistry, where it is typically also
reflected by the appearance of an induction period \cite%
{Nicolis,PolymerScience1,PolymerScience2,PolymerScience3,PolymerScience4}.
Hierarchical self-assembly is also ubiquitous in biology. In particular, in a series of papers on the polymerization and crystallization
of sickle cell Hemoglobin (HbS) --an issue of crucial importance for the
pathophysiology of sickle cell anemia-- it has been shown that metastable
dense liquid clusters serve as precursors to the ordered nuclei of the HbS
polymerization\cite{Vek1,Vek2,Vek3}. Induction times causing time delay have
been observed and analyzed during this non-standard nucleation process of
the HbS polymers. More generally, it is now known that homogeneous
nucleation -most notably in nanophase materials such as proteins but
apparently even in simple fluids - can involve a two step process starting
with the formation of more structured liquid-like droplets and/or clusters %
\cite{Vek3, Oxtoby1, Oxtoby2} from the solution followed by the development
of crystalline order \cite{MST2008,Lutsko}. Here, the material in solution
can be viewed as the population N of our scheme, the dense-liquid phase as
the intermediate, X, and the solid as S. Cooperativity is again responsible
for the induction period and time delays during the subsequent ordering
corresponding to the processes $N \rightarrow X$ and $X \rightarrow S$.

A class of nanophase materials in which self-assembly appears to be
accompanied by a rich variety of unexpected behaviors are synthetic zeolites~%
\cite{MST2008, JJK07, Barrer82, Kirschhock2005, Kremer03a,Kremer03b} which
find nowadays applications in a broad range of areas such as heterogeneous
catalysis, petroleum refining, and microelectronics~\cite{clerici00,
eslava07}. In the context of zeolite formation mechanism, Silicalite-1 type
zeolite~\cite{FLANIGEN78} in particular has been the subject of numerous
studies~\cite{Watson97,Schoeman97zeo,Schoeman97mic,deMoor99b,
deMoor2000,Mintova02,Houssin03,fedeyko04,Rimer05JPCB,Pelster06,
Davis2006,Patis07,tom1,tom2,Provis2008,AertsNMR}.

\begin{figure}[tbp]
\begin{center}
\resizebox{8cm}{!}{
{
\includegraphics[width = 8cm, angle=0]{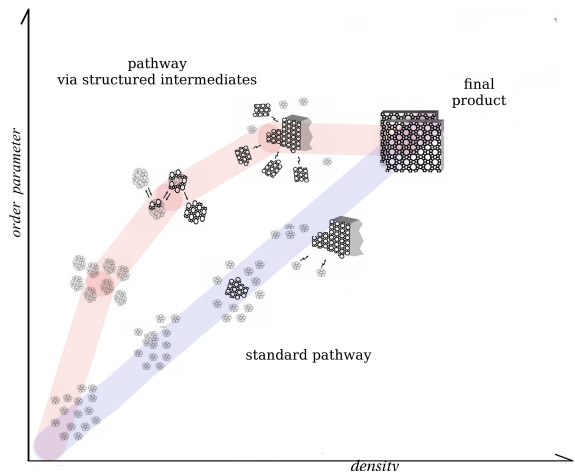}
}}
\end{center}
\caption{ A schematic demonstration of multistep versus standard synthetic
pathways in a two-order parameter space accounting for the presence of
partially structured intermediates. }
\label{fig1}
\end{figure}

\begin{figure}[tbp]
\begin{center}
\resizebox{9.0cm}{!}{
{\includegraphics[width = 9.0cm, angle=-0]{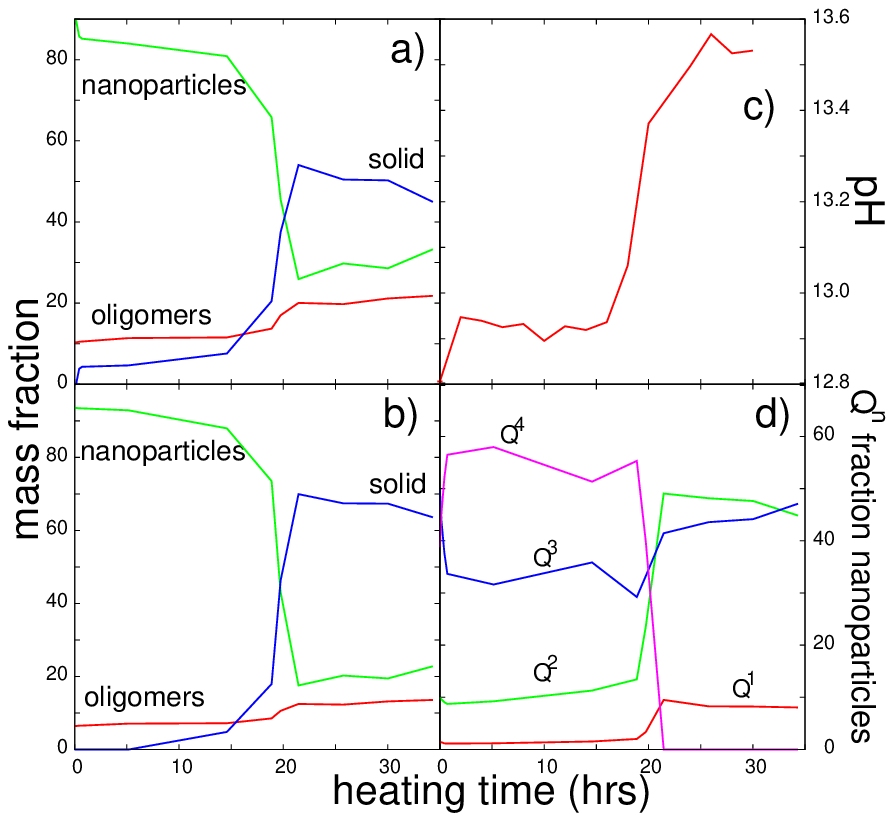}}}
\end{center}
\caption{(Color on line) Liquid state $^{29}$Si NMR and pH results from
Ref.~[\onlinecite{AertsNMR}]. (ab) Evolution of nanoparticles (N and similarly sized X particles), oligomers
and NMR- undetected phase (denoted as "solid"). The undetected phase
consists of large particles, i.e. crystalline solid ($S$) and possibly a
fraction of intermediates (X particles that are considerably larger than nanoparticles), that fall outside the NMR detection
limit. The oligomers are an additional group of smaller silicate species,
but their presence is assumed to be of only secondary importance for the
crystal growth. Panel-a: direct NMR results obtained with a
short delay time of 7 sec between radiofrequency pulses in the NMR
experiment. The process shows two-step evolution. The first
step ($<$ 1hr) indicates the formation of intermediate species.
Subsequently, there was a long induction period during which the amount of
nanoparticles varied slowly, followed by a rather abrupt transition towards
an equilibrium state dominated by zeolite crystalline material. It is noted
that in the original publication, these NMR data were corrected to account
for a longer NMR delay time of 92 s (Panel-b), based on measurement of the
unheated sample~\protect\cite{AertsNMR}. The correction resulted in
quantitatively more reliable populations but the initial step was
suppressed. The two-step evolution was also evident from measurements of the
suspension pH (panel-c) and the connectivity of the nanoparticle framework
(panel-d). The connectivity is expressed as $Q^n$ fractions, where $Q^n$
stands for a silicon atom connected to $n = 1-4$ neighboring silicons via
siloxane bonds. The disappearance of the $Q^4$ fraction might indicate a depletion of the more structured X-particles.}
\label{fig2}
\end{figure}

The starting point of Silicalite-1 synthesis is a suspension of ~3 nm
silicate nanoparticles~\cite{Schoeman97mic}. The first period of ageing, at
room or elevated temperature, is characterized by a rapid evolution of these
nanoparticles. During this period, the average nanoparticle diameter
increases to a slightly larger value of 5-6 nm~\cite%
{Schoeman97zeo,Rimer05JPCB,Davis2006,AertsNMR}. On a structural level, the
nanoparticle silicate network condenses, i.e. additional siloxane (Si-O-Si)
bonds are formed~\cite{Patis07, Provis2008,AertsNMR}. Subsequently, still
larger particles start growing, ultimately resulting in Silicalite-1 zeolite
crystals. Convincing evidence shows that Silicalite-1 crystals grow by
aggregation of the 5-6 nm nanoparticles~\cite{Davis2006}. It was proposed
that a fraction of the condensed 5-6 nm nanoparticles develops a zeolite
crystalline framework. These intermediate nanoparticles can be identified as
"nuclei" for zeolite crystal growth by nanoparticle aggregation~\cite%
{Davis2006}. The picture is then one of nanoparticles N giving rise to
intermediates X which, in turn, form the (zeolite-crystalline) solid S, as
illustrated in Fig.~\ref{fig1} Recent experimental liquid state $^{29}$Si
NMR and pH results~\cite{AertsNMR} of Silicalite-1 zeolite crystallization
are presented and described in Fig~\ref{fig2}.

Fig~\ref{fig2}-ab shows the experimental distribution of silicon among
different species in suspension, determined with liquid state $^{29}$Si NMR.
There is a long induction period during which the amount of nanoparticles
varies slowly, followed by a rather abrupt transition towards an equilibrium
state dominated by zeolite crystalline material (solid). Because
nanoparticles N and intermediates X have similar size NMR cannot distinguish
between these two species. The rapid condensation of nanoparticle framework
in the initial stage, resulting in the formation of intermediates, is
evident from the silicon connectivity distribution (Fig~\ref{fig2}-d) and
the corresponding increase of the suspension pH (Fig~\ref{fig2}-c). Upon
transition to crystalline material, the pH shows a second increase.

Notice that during the later stages of crystallization, other mechanisms,
such as Ostwald ripening, acting on the already formed aggregates are likely
to take over as dominant crystallization mechanism.

In view of the foregoing we argue that a timescale separation and the
appearance of long quasi-stationary plateaus is likely to be a generic
property of large classes of synthetic processes. One of the main goals of
this article is the theoretical explanation of the mechanisms underlying
these phenomena. We establish a generic theoretical model which captures the
main features of self-organization in the process of multi-step nucleation
and growth. Specifically, we show how the combination of
intermediate steps in self-assembly, together with first-order nucleation
and cooperativity, in the form of growth controled by the available surface
area of the product, can explain the long induction times often
observed in these processes. For the
particular case of zeolites, our model is intended to sort out the
basic microscopic mechanisms that control the nucleation and growth of the final
crystalline product. This can have some practical repercussions in not only eliminating superfluous conjectures that could be entertained in the absence of theoretical input, but also in further designing targeted and better controlled experiments.

The article is organized as follows: In Section 2, we outline the proposed
mechanisms for self-organization incorporated in our mathematical models. In
Section 3 we develop a simple phenomenological model based on the assumption
of cooperative behavior which reproduces much of the qualitative behavior
summarized above. A more microscopic and mechanistic model is presented in
Section 4 and shown to describe the generic qualitative features of recent
observations and lead to a prediction regarding the distribution of cluster
sizes. The main conclusions are summarized in Section 5.

\section{The mechanisms for self-organization and their modeling}

\label{secmech}

In this section we outline the construction of mathematical models capable
of quantifying the mechanisms summarized in the Introduction and, in
particular, the presence of multiple time scales and the existence of a long
induction period. The models to be developed are articulated around three
basic steps which account for the self-organizing behavior.

\begin{description}
\item[(a)] \emph{Equilibration between nanoparticles and intermediates. }
This part of the dynamics will depend on the details of specific systems.
Here, we will model it as a globally autocatalytic process whereby nanoparticles and
intermediates interact to form new intermediates while depleting the
nanoparticle population. The rationale behind this
assumption is that in the particular case of zeolites, the intermediate is
not a single species but, rather, it is believed to consist of a population
of structures of various sizes\cite{deMoor99b,AertsNMR}. We assume that the dominant mechanism
for conversion of nanoparticles into intermediates is growth of smaller
intermediates via aggregation of nanoparticles, i.e., via an autocatalytic
mechanism. It would be possible to include nucleation of new intermediates,
but it seems likely that growth will be the dominant process so that, for
simplicity, nucleation can be neglected. It is assumed that at the
beginning, a small population of intermediates is already present due, e.g.,
to thermal fluctuations. Obviously, for other systems, such
as proteins, first-order reactions characteristic of nucleation would be
more appropriate.

\item[(b)] \emph{Initial formation of final product. } At some point, a
stable cluster of S must be formed which will then proceed to grow and we
assume that the long plateau referred to earlier is a measure of the time
required for this to occur. In a simple fluid this would be recognized as a
nucleation phenomena\cite{Kashchiev} which is typically modeled as a
probabilistic process occurring with a probability per unit time
proportional to the supersaturation. In the case of more complex materials,
it may be that the intermediates must pass through multiple structural
phases or that some other type of transformation of the intermediates is
necessary before stable clusters of final product can form. It might be more
appropriate in this case to speak of self-assembly rather than nucleation%
\cite{Jensen}. Nevertheless, we will assume that this complex microscopic
process can be crudely described in the same terms as nucleation in a simple
fluid.

\item[(c)] \emph{Growth and Cooperativity. } Once the product begins to
form, the speed of the process increases dramatically. This is a signal of
cooperative behavior whereby the presence of some product accelerates the
formation of new product. One possibility is that this is simply due to the
growth of the clusters once they are formed. In simple fluids, secondary
nucleation - the nucleation of new crystals on the surfaces of existing
crystals - also occurs and can speed the formation of crystals\cite{Mullin}.
In more complex materials, if the intermediates must pass through different
phases, then a growing cluster could also catalyze that process thereby
accelerating cluster formation. We will not try to distinguish these
different mechanisms but will simply account for an acceleration of the rate
of cluster formation due to the presence of existing product.
\end{description}

We will discuss the implementation of this phenomenological framework in two
stages. First, we translate it into a crude phenomenological model which
serves to illustrate the general ideas. Then, we describe a more
microscopic, mechanistic approach that leads to similar results and that has
the advantage of being less ad hoc and of making further predictions.

\section{Phenomenological Model}

\label{secphen}

We consider a closed reactor initially containing nanoparticles, N, present
in abundance, along with an intermediate (more ordered and larger) species,
X, present in small amounts. Rather than forming the product, S, directly,
the N particles first accumulate to form the intermediate X. In fact, the single interemediate in the model represents a spectrum of species. These species undergo reactions which convert from one type of intermediate to another, but this will not be modeled here. Instead, we only track the net mass of intermediates which increases by consuming nanoparticles. In certain cases, such as zeolite crystallization, it is believed that the intermediates are the only species possessing the necessary structure that allows them to bind together to form solid\cite{AertsNMR}. For this reason, we restrict our analysis to the sequence $N \rightarrow X \rightarrow S$ and do not consider possibilities such as direct conversion of nanoparticules into solid which may be of relevance to other processes. Such alternatives could easily be included within the framework described below. The N to X
transition is taken to be an \emph{autocatalytic} process of order higher
than one, with a rate $\nu_{1}$ depending on the abundance of both N and X.
Furthermore, in agreement with available data it is stipulated that the X particles accumulate to give rise to S. The X to S
transition occurs at a rate, $\nu_{2}$ , that describes two processes.
First, as discussed above, the initially slow formation of stable clusters
is modeled as a probabilistic event occurring at a rate that depends on the
supersaturation. Then, as the final product is formed, the rate increases
due to cooperativity. Thus, the total rate of formation of the product will
depend on the concentration of intermediates and of the product itself. To
avoid the presence of stoichiometric coefficients we choose to work with the
mass fractions $n$, $x $ and $s$ of N, X and S particles respectively since
by mass conservation the rate of loss of $n$ in the N to X transition will
then necessarily be equal to the rate of growth of $x$, etc. We obtain in
this way the following three coupled differential equations, 
\begin{eqnarray}
\frac{dn}{dt} & = & -\nu_{1}(n,x )  \notag \\
\frac{dx}{dt} & = & \nu_{1}(n,x )-\nu_{2}(x,s )  \notag \\
\frac{ds}{dt} & = & \nu_{2}(x,s ),  \label{eq:CORE}
\end{eqnarray}
the corresponding kinetic scheme being 
\begin{eqnarray}
N\rightleftharpoons X & & (\text{ rate}\;\nu_{1})  \notag \\
X\longrightarrow S & & (\text{ rate }\: \nu_{2})  \label{eq:NtoXtoS}
\end{eqnarray}
Notice the conservation condition 
\begin{equation}
n+x+s=constant  \label{eq:MASS}
\end{equation}
This relation does not involve the mass fraction of
oligomers as they are believed to play a secondary role in the process and
are therefore neglected\cite{AertsNMR}.

To illustrate this behavior, we choose to model the rate $\nu_{1}$ as a
Verhulst type growth process\cite{LogisticMap,Nicolis}, 
\begin{equation}
\nu_{1}(n,x )=k_{1}nx-k_{2}x^{2}  \label{eq:Verhulst}
\end{equation}
whereby nanoparticles and intermediates react to produce more intermediate
species. Notice that balance is achieved when $n=\frac{k_2}{k_1}x$ so that
the ratio of rate constants can be fixed by the quasi-equilibrium level of
the mass fractions which occur at short times. The value of, say $k_1$ is
then determined by the time the system takes to reach this quasi-stationary
state. Thus, these parameters are completely determined by the short-time
behavior of the system.

The rate $\nu_{2}$ is broken into two parts: $\nu_2(x,s )=\nu_{21}(x
)+\nu_{22}(x,s )$. The first part depends only on the amount of X present
and represents the nucleation/self-assembly process. For this, we take the
simplest reasonable form, $\nu_{21}(x )=k_{3} (x-x_0)$, where $x_0 $ is the
equilibrium concentration of X so that $x-x_0$ represents the
supersaturation. The second part of the rate represents the effect of
cooperativity. We have not found it possible to model the
behavior observed in the zeolite data using higher order terms involving
only the concentration of intermediates. Instead, we have found it necessary to model this as an
enhancement of the nucleation rate that depends on the amount of product
already formed, $\nu_{22}(x,s )=k_{3}\alpha s^{\mu}(x-x_0)$. In the
following we always assume that $x_0=0$ which implies that once the product
is formed from X, it will not dissolve again.

Putting these pieces together, the phenomenological model is given by 
\begin{eqnarray}
\frac{dn}{dt} & = & -k_{1}nx+k_{2}x^{2}  \label{pmodel} \\
\frac{dx}{dt} & = & k_{1}nx-k_{2}x^{2}-k_{3} x(1+\alpha s^{\mu})  \notag \\
\frac{ds}{dt} & = & k_{3}x(1+\alpha s^{\mu})  \label{eq:NXS-nucl}
\end{eqnarray}
Long induction times are expected to arise when the cooperative part of the
process is slow, which is to say that $k_{3} << k_{1},k_{2}$. In this case,
the model exhibits dynamics on two different time scales: at short times,
the cooperative part can be neglected with the result that $c=n+x$ is
conserved (since virtually no product forms) and the model can be integrated
to give $n(t)=c-x(t)$ and 
\begin{equation}
x(t)=\frac{k_{1}cx(0)}{k_{1}c-(k_{1}+k_{2})x(0)\left( e^{-k_{1}ct}-1 \right)}
\end{equation}
For $k_{1}ct >> 1$, this leads to an equilibrium with $x=\frac{k_{1}c}{%
k_{1}+k_{2}}$. Thus, assuming that $c$ is on the order of $1$, we find
equilibration on a time scale determined by $k_{1}$ and with relative
fractions of material in the form of N and X determined by $k_{2}$. This
short-time equilibration is expected to describe the beginning of the
self-assembly process when materials are first mixed, heat is first applied,
etc.

\begin{figure}[tbp]
\begin{center}
\resizebox{8cm}{!}{
{\includegraphics[width = 8cm, angle=-90]{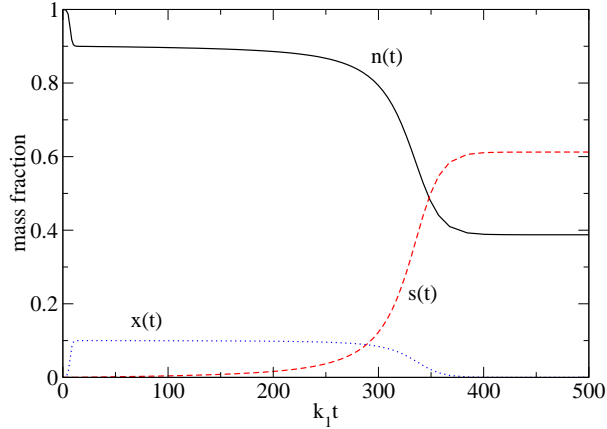}}}
\end{center}
\caption{(Color on line) Solution of the phenomenological model, Eq.(\ref%
{pmodel}), with the rate constants given in the text. The full line is the
mass fraction of the nanoparticles, the dotted line is that of the
intermediates, $x(t)$, and the broken line is that of the final product, $%
s(t)$}
\label{fig3}
\end{figure}

At longer times, nucleation and cooperativity contribute causing the system
to evolve away from the short-time steady state. There is then a new,
long-time steady state in which $x \rightarrow 0$. At intermediate times,
the model must be solved numerically. Figure \ref{fig3} shows the result of
such a solution starting with initial conditions $x(0) = 10^{-4}$ and $n(0)
= 1-x(0)$, corresponding to an initial state in which almost all the mass is
in the form of nanoparticles. We take $k_{2} = 9k_{1}$ so that the
short-time equilibration leads to $n=0.9$ and $x=0.1$. To illustrate the
presence of a long induction time, the value $k_{3} = 0.002k_{1}$ was used
so that if $1/k_{1}$ is on the order of a minute, then significant
self-assembly of the final product only occurs after a time on the order of $%
1/k_{3}$ or about $8$ hours. The remaining constants, $\alpha$ and $\mu$,
primarily control the speed with which the system makes the transition from
its short-time steady state to the final long-time steady state. These
values were chosen somewhat arbitrarily to be $k_{3} \alpha = k_{1}$ and $%
\mu = 1.75$. The figure shows that after a rapid equilibration, the system
experiences a long period of slow nucleation that is followed by a rapid
transition to the final steady state. The final transition begins to occur
around $k_{1}t \sim 250$ correspond to $t \sim 4$hr for $k_{1}=1/$minute.
Qualitatively, this behavior is similar to that observed in the zeolite
experiments of Aerts et al\cite{AertsNMR} (compare to Fig. \ref{fig2}).

Finally, we note some specific features of this model that
are motivated by the experimental data shown in Fig. \ref{fig2}. First,
there is no evidence of nucleation of intermediates from nanoparticles. This is in part
based on the fact that the nanoparticle concentration reaches a finite
plateau even though the intermediates are depleted. Nucleation may a priori occur on much longer timescales than that of the
experiment in which case the observed plateau is only quasi-stationary. This
would then suggest adding a term $-k_{4}n$ to the equation for $dn/dt$ and
a corresponding term in the equation for $dx/dt$. However, the rate constant
would be very small thus justifying our neglect of this term on these
timescales. Second, the conversion of intermediates to solid is
irreversible. This is motivated by the fact that dissolution of zeolite crystals in alkaline medium does not lead to the extraction of silicate species of the size of nanoparticles or larger\cite{Groen} (which also explains
the depletion of X (see fig. 2d)). This leaves open the possibility of an equilibration between the solid and the oligomers, but, as can be seen from Fig. \ref{fig2}, such a reaction, if it exists, is of secondary importance and must take place on longer time scales than that of the formation of solid from nanoparticles.  Besides, the fact that the crystals eventually grow to macroscopic size suggests that the formation of solid is, for all practical purposes, irreversible.

\section{A more microscopic approach}

\label{microscopic}

Having shown that the combination of nucleation/self-assembly and
cooperativity appears to give a reasonable description of the mechanisms of
self-assembly, we now describe a more concrete realization of this picture.
It is expected that assumptions made here are both more physically plausible
than those made above and more intuitively appealing. They are:

\begin{enumerate}
\item The $N\rightleftharpoons X$ reaction is the same as assumed in the
previous section.

\item Formation of new stable clusters of final product occurs at a rate
proportional to the supersaturation.

\item The initial size of a stable cluster is always the same.

\item Once formed, clusters grow at a rate proportional to their surface
area and to the supersaturation.

\item All clusters are geometrically identical.

\item The formation rate is enhanced by a factor proportional to the surface
area of the existing clusters and to the supersaturation.
\end{enumerate}

Assumptions 1 and 2 are the same as made previously. Assumption 3 is perhaps
the most problematic: in a simple fluid it might be justified on the grounds
that growing clusters are presumed to start out as a stable cluster which
has a fixed size (at least for a given supersaturation). This may be true
here. If not, it can be imagined that there is a distribution of initial
cluster sizes and that we are describing an average over an ensemble of such
clusters. Assumption 4 explicitly introduces cluster growth which was
neglected in the previous section (or, at least, it was crudely lumped
together with cooperativity). Assumption 5 is merely a convenience so that
we can be specific concerning geometric factors such as the ratio of volume
to surface area of a cluster. Assumption 6 seems the simplest, physically
plausible way to introduce the idea of cooperativity.

These models involve the area, $A$,  and mass or,
equivalently, volume, $V$,  of growing clusters. For spherical
clusters, we of course expect that $A\sim R^{2},V\sim R^{3}$ where 
$R$ is the radius. In fact, these scaling relations should hold
for three dimensional growth, with $R$ some measure of linear
size,  regardless of the precise shape of the clusters. However, other modes
are possible and are of relevance for growth of some protein molecules. For
example, suppose that the cluster has the shape of a cylinder with
crossectional area $\pi r^{2}$ and length $L$.
Quasi-one-dimensional growth would occur if the cylinder lengthens with
constant cross section. This would involve attachment on the flat faces so
that the active growth area would be $2\pi r^{2}$, a constant,
while the volume would be $\pi r^{2}L(t)$. Quasi-two-dimensional
growth would occur if attachment happened on the curved part of the cylinder
so that the crossectional area increases with time while the length remains
constant. Then, the active growth area would be $2\pi r\left( t \right) L$ and the volume $\pi r^{2}\left( t\right) L$. All of these
cases can be summarized by the general scaling $A=G_{A}R^{D-1},V=G_{V}R^{D}$ where $R$ is the appropriate linear length scale and $G_{A}$ and $G_{V}$ are constant geometric factors. We
will use this as our general model with the additional provision that $D$ is an integer. We therefore rule out fractal growth patterns which
could a priori be treated within our general framework but which do not admit of the
mathematical simplifications we utilize below.

To translate this into a mathematical model for growth, some new quantities
must be defined. Let $\mathcal{N}(t)$ be the number of clusters growing at
time $t$ and let $m(t,t^{\prime })$ be the mass, at time $t$, of a cluster
nucleated at time $t^{\prime }$. Note that by assumption 3, $m(t,t)=m_{0}$
for some constant $m_{0}$. Finally, the total mass of nanoparticles,
intermediates, solid and oligomers is denoted $M$, and is a constant
throughout the experiment. The total mass of solid at time $t$, $Ms(t)$, is
simply given by a sum over the masses of all clusters present at time $t$. A
cluster at time $t$ has mass $m(t,t^{\prime })$ if it was nucleated at time $%
t^{\prime }$. The number nucleated at time $t^{\prime }$ is simply $\frac{d%
\mathcal{N}(t^{\prime })}{dt^{\prime }}dt^{\prime }$ so, 
\begin{equation}
s(t)=M^{-1}\int_{0}^{t}m(t,t^{\prime })\frac{d\mathcal{N}(t^{\prime })}{%
dt^{\prime }}dt^{\prime }  \label{stM}
\end{equation}%
where we assume that there is no final product present at the beginning of
the experiment, $s(0)=0$. Thus, based on assumption 4, the mass of a cluster
grows as 
\begin{align}
\frac{dm(t,t^{\prime })}{dt}& =k_{g}xA(t,t^{\prime })  \notag \\
m(t,t)& =m_{0}
\end{align}%
for some constant, $k_{g}$. Finally, the rate of appearance of new clusters
is the sum of a nucleation-like term proportional to the supersaturation
(assumption 2) and a cooperative term proportional to the total area of the
clusters (assumption 6), 
\begin{equation}
\frac{d\mathcal{N}(t)}{dt}=Mk_{n}x+k_{a}x\int_{0}^{t}A(t,t^{\prime })\frac{d%
\mathcal{N}(t^{\prime })}{dt^{\prime }}dt^{\prime }  \label{NMk}
\end{equation}%
Note that we include a factor of the total mass in the nucleation term. This
is because the number of clusters is an extensive quantity - as is the total
area calculated in the second term. The number of nuclei generated per unit
time must also be extensive, hence the factor of the mass.

At this point, the model seems much more complicated than the
phenomenological model given above. However, after some simple manipulations
(see Appendix), it can be described by a set of three first order
differential equations which uses the same number of parameters as in the
phenomenological model: 
\begin{eqnarray}
\frac{dn}{dt} &=&-k_{1}nx+k_{2}x^{2}  \label{model} \\
\frac{dx}{dt} &=&k_{1}nx-k_{2}x^{2}-\gamma x\frac{d\widetilde{s}(u)}{du} 
\notag \\
\frac{du}{dt} &=&\gamma x  \notag
\end{eqnarray}%
Here, $u(t)$ is an auxiliary function with the initial condition $u(0)=0$.
The mass fraction of $S$ is given by $s(t)=\widetilde{s}\left( u(t)\right) $
where%
\begin{equation}
\widetilde{s}\left( u\right) =\left( 1+\frac{D}{\alpha }\right)
\sum_{i=1}^{D}a_{i}\exp \left( \lambda _{i}u\right) -D\frac{\beta }{\alpha }u
\label{su}
\end{equation}%
The constants $\lambda _{i}$ are the roots of the $D$-th order equation%
\begin{equation}
0=\lambda ^{D}-\alpha \sum_{j=1}^{D}\frac{\left( D-1\right) !}{\left(
D-j\right) !}\lambda ^{D-j}
\end{equation}%
while the coefficients $a_{i}$ are%
\begin{eqnarray}
a_{i} &=&\beta \frac{1}{2\lambda _{i}-\lambda _{1}-\lambda _{2}},\;D=2 \\
a_{i} &=&\beta \frac{\alpha -\sum_{j\neq i}\lambda _{j}}{\prod\limits_{j\neq
i}\left( \lambda _{i}-\lambda _{j}\right) },\;D=3  \notag
\end{eqnarray}%
The case of general values of $D$ is discussed in the Appendix. All of this
is written in terms of the two dimensionless constants%
\begin{equation}
\alpha =\frac{Dk_{a}m_{0}}{k_{g}},\quad \beta =\frac{m_{0}DG_{V}\rho
R_{0}k_{n}}{G_{A}k_{g}}  \label{defab}
\end{equation}%
and the rate constant 
\begin{equation}
\gamma =\frac{k_{g}G_{A}}{DR_{0}\rho G_{v}}  \label{defgamma}
\end{equation}%
where $m_{0}=\rho G_{V}R_{0}^{D}$. Thus, despite the fact that the physical
model involves 3 parameters, $k_{g}$,$k_{n}$ and $k_{a}$, as well as the
initial mass of a stable cluster, $m_{0}$, and the total mass $M$ and the
density $\rho $, the data can be fit with only three parameters, $\alpha
,\beta $ and $\gamma $, which is no more than the phenomenological model.

The present model is now written in terms of rate equations which are
exactly equivalent to eqns~(\ref{stM}-\ref{NMk}). Still, it does not have
the same form as the phenomenological model proposed in the previous
Section. It is, however, possible to make a correspondence between them.
Since both the mass fraction of $S$, $s(t)$, and the auxiliary variable $%
u(t) $ increase monotonically with time, and since they are related by a
simple algebraic equation, eqn~(\ref{su}), we can imagine inverting that
equation to get $u$ as a function of $s$. An analytic inversion is not
possible but for small $u$ and $s$, this can be done perturbatively giving $%
u $ as a power series in $s$. (See the Appendix for details.) Substitution
into eqn~(\ref{model}) and specializing to $D=3$ gives 
\begin{align}
\frac{dn}{dt}& =-k_{1}nx+k_{2}x^{2}  \label{model-rate} \\
\frac{dx}{dt}& =k_{1}nx-k_{2}x^{2}-\beta \gamma x\left( 1+\frac{\alpha +3}{%
\beta }s-\frac{\alpha +3}{2\beta ^{2}}s^{2}+...\right)  \notag \\
\frac{ds}{dt}& =\beta \gamma x\left( 1+\frac{\alpha +3}{\beta }s-\frac{%
\alpha +3}{2\beta ^{2}}s^{2}+\ldots \right)  \notag
\end{align}%
which is very similar to the phenomenological model except that the single
term $s^{\mu }$ is replaced by an infinite series in $s$. Thus the
adjustment of $\mu $ is seen as a way of accounting for the fact that the
infinite sum is being approximated by a single term. 
\begin{figure}[tbp]
\begin{center}
\resizebox{8cm}{!}{
{\includegraphics[width = 8cm, angle=-90]{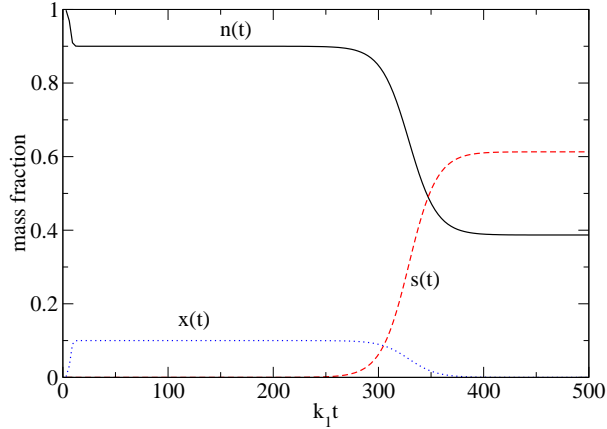}}}
\end{center}
\caption{(Color on line) Solution of the microscopic model, Eq.(\ref{model}%
), (lines) with the rate constants given in the text. The full line is the
mass fraction of the nanoparticles, the dotted line is that of the
intermediates , $x(t)$, and the dashed line is the mass fraction of the
solid, $s(t)$.}
\label{fig4}
\end{figure}

Figure \ref{fig4} shows the result of a numerical solution of the model, Eq.(%
\ref{model}). The constants $k_{1}$ and $k_{2}$ have the same values as used
previously. The other constants, $\alpha =8750$, $\beta =1.67\times 10^{-3}$
and $\gamma =6.55\times 10^{-5}k_{1}$, were chosen to give approximately the
same point of crossing of the curves and similar final states as seen in the
phenomenological model. In this case, the plateau resulting from the long
induction time is even more pronounced and the transition to the steady
state is even sharper than in the phenomenological model.

\begin{figure}[tbp]
\begin{center}
\resizebox{8cm}{!}{
{\includegraphics[width = 8cm, angle=-90]{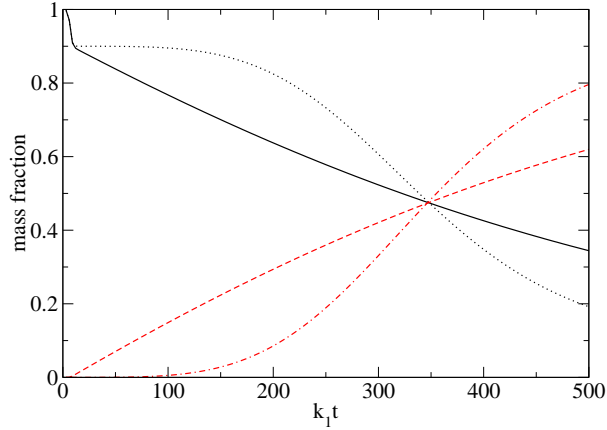}}}
\end{center}
\caption{(Color on line) Comparison of the model without cooperativity and
the data. The solid and dotted lines are the nanoparticle mass fractions and
the remaining curves are the solid mass fraction. The solid and dashed lines
were calculated using $\protect\beta =5.0$ and $\protect\gamma %
=0.0.0034k_{1} $; the dotted and dash-dotted lines correspond to $\protect%
\beta =5\times 10^{-6}$ and $\protect\gamma =2.66k_{1}$. Decreasing $\protect%
\beta $ gives a slightly longer plateau but never as long and sharp as that
possible when cooperativity is present.}
\label{fig5}
\end{figure}

To test whether or not cooperativity is really required, we have attempted
to model similar behavior with $k_{a}=0$. Typical results are shown in Fig. %
\ref{fig5}. In this case, there are only two parameters and we find that it
is not possible to simultaneously capture the long plateau up to $k_{1}t
\sim 200$ and to capture the crossing of the curves at about $k_{1}t \sim
310 $. Furthermore, in contrast to both the data and the model with
cooperativity, the mass fractions do not show plateaus at long times. As can
be seen in the figure, a decrease in $\beta $ by a factor of 
 $10^{-6}$, with $\gamma $ adjusted to
fix the point at which the curves cross, has only a small effect on the
quality of the fit to the data. Further decrease in $\beta$
has little  effect on the length of the plateau. We conclude that
the cooperativity is required to explain the presence of a long induction
time.

One advantage of the microscopic approach is that, as discussed in the
Appendix, the size distribution of the clusters is easily calculated. The
number of clusters with radius $R$ or smaller at time $t$, $\mathcal{N}%
\left( R,t\right) $, is given by%
\begin{equation*}
\mathcal{N}\left( R,t\right) =\left\{ 
\begin{array}{ll}
0, & \text{ if }R\leq R_{0} \\ 
\frac{M}{m_{0}}\big(\widetilde{\mathcal{N}}\left( u(t)\right) & -\widetilde{%
\mathcal{N}}\left( u\left( t\right) +1-\frac{R}{R_{0}}\right) \big), \\ 
& \text{ if }R_{0}<R<R_{0}\left( u(t)+1\right) \\ 
\frac{M}{m_{0}}\widetilde{\mathcal{N}}\left( u(t)\right) , & \text{ if }%
R\geq \left( u(t)+1\right) R_{0}%
\end{array}%
\right.
\end{equation*}%
The function $\widetilde{\mathcal{N}}\left( u\right) $ is related to the
total number of clusters by $\mathcal{N}\left( t\right) =\frac{M}{m_{0}}%
\widetilde{\mathcal{N}}\left( u(t)\right) $ and is closely related to $%
\widetilde{s}(u)$. Its explicit form is%
\begin{equation}
\widetilde{\mathcal{N}}(u)=\sum_{j=1}^{D}a_{j}\exp \left( \lambda
_{j}u\right) .
\end{equation}%
The factor $R_{0}\left( u\left( t\right) +1\right) $ occurring in the size
distribution is the radius of the largest possible cluster at time $t$ and
the prefactor $\frac{M}{m_{0}}$ is the maximum number of clusters that could
be created from the given amount of material.

\section{Conclusion}

\label{seccon}

We have shown that a self-organization process involving
self-assembly/nucleation, growth and cooperativity mediated by intermediates
consisting of a polydisperse population of nanosized particles with varying
structure accounts for the kinetics of self-assembly observed in a class of
nano-sized materials. In particular, the models reproduce a long
quasi-steady state and the long waiting times observed as well as the final
apparent equilibrium between nanoparticles and clusters of final product.
Note that the reason for the termination of the self-assembly process in our
model is due to the depletion of the particles, X. At some point, the $N
\rightarrow X$ transitions can not compensate for the fast incorporation of
X into the final phase. As new particles can only be formed when some X is
already present, the whole process stops when X is fully consumed. The model
that we have introduced may account for a broad class of aggregation
processes. There is evidence that the complex mechanism of zeolite
synthesis, at least some features of it, can be described by this model. The
two-step behavior of Fig.~\ref{fig2}-a resembles the evolution of article
populations as depicted in Figs.~\ref{fig3} and \ref{fig4}, especially if we
assume that some part of the (larger) intermediates X belong to the
undetected (solid) phase. Moreover, the extinction of structured
nanoparticles/intermediates which announces the end of the growth process,
seems confirmed by experiment (compare $Q^4$ in Fig.~\ref{fig2}-d with $X(t)$
in Figs.~\ref{fig3} and \ref{fig4}).

The basic physical mechanisms were modeled at two levels. First, a purely
phenomenological model was presented which captures the basic ideas in their
simplest form. Then, a more detailed and mechanistic picture was given. This
involved new concepts, such as the distribution of cluster sizes, but in the
end was reduced to a system of rate equations similar to the first model. In
fact, it was noted that the phenomenological model could be viewed as a
systematic approximation to the more detailed model.

It was shown that the concept of cooperativity plays a key role in
explaining the observations. When cooperativity is removed from the model,
it is not possible to simultaneously reproduce the long waiting time at the
beginning of the process and the rapid growth that occurs after long times.

One point that deserves emphasis is the limited nature of the assumptions
underlying our model. It is based on analogies to the familiar ideas of
crystallization\cite{Mullin} but we do not claim that the processes are
necessarily the same. The nucleation event that initializes crystallization
in a simple fluid is probably a much simpler process than the self-assembly
in more complex materials such as zeolites. However, it makes sense that in
both cases the rate at which the process happens must surely, in a first
approximation, be proportional to the supersaturation. Similarly, the
mechanism of cooperativity in a simple fluid is simply the accelerating
growth of crystallites as they become larger. In some cases, secondary
nucleation on growing crystals is also a factor. In zeolites, however, it
may be a true catalysis of the transformation of intermediaries from one
form to another. Whatever the nanoscale mechanism, it is plausible that in a
first approximation the rate must be proportional to the surface area of
existing clusters and to the supersaturation. Our results strongly suggest
that \emph{some} form of cooperativity, whose specifics will depend on the
system at hand, is necessary when long induction times precede rapid
transformation from one state to another.

It would be interesting to compare the results of our model with other
experimental data~\cite{AertsNMR} and theoretical models that have been
proposed~\cite{Drews07} for zeolites. Moreover, it is hoped that this model
will provide a framework for pursuing further experiments. In particular,
the model makes a definite prediction concerning the distribution of cluster
sizes and it would seem that this could be checked via experiment.

\begin{acknowledgments}
This work was supported in part by the Prodex programme of the European
Space Agency under contract number C90241. CEAK, TSvE, and JAM acknowledge
the Flemish government for long-term structural support via the centre of
excellence (CECAT), the concerted research action (GOA), and Methusalem
funding. TPC and AA were supported by an IWT and FWO grant respectively.
\end{acknowledgments}

\bigskip \bigskip

\renewcommand{\theequation}{A.\arabic{equation}}

\appendix{}

\section{Elaboration of the model of Section \ref{microscopic}}

\subsection{Derivation of eqns~(\ref{model}-\ref{defgamma})}

The model for the formation of solid can be summarized as%
\begin{eqnarray}
s(t) &=&M^{-1}\int_{0}^{t}m(t,t^{\prime })\frac{d\mathcal{N}(t^{\prime })}{%
dt^{\prime }}dt^{\prime }  \notag \\
\frac{dm(t,t^{\prime })}{dt} &=&k_{g}xA(t,t^{\prime })  \notag \\
\frac{d\mathcal{N}(t)}{dt} &=&Mk_{n}x+k_{a}x\int_{0}^{t}A(t,t^{\prime })%
\frac{d\mathcal{N}(t^{\prime })}{dt^{\prime }}dt^{\prime }
\end{eqnarray}%
The mass and surface area of a growing cluster will be related in a
system-specific way. For example, if the cluster is spherical, then the area
will scale as the square of the radius and the mass as its cube. On the
other hand, if the growth is primarily one-dimension, e.g. a cylinder which
grows by becoming longer, then the mass and area will both increase in time
as the first power of the length. We will assume in general that there is
some length scale, $R(t,t^{\prime })$ characterizing the size of clusters at
time $t^{\prime }$ which have nucleated at time $t$ and that the area scales
as $A(t,t^{\prime })=G_{A}R^{D-1}(t,t^{\prime })$, where $G_{A}$ is a
geometric factor and $D$ is the dimensionality of the growth process so that
the mass will scale as $m(t,t^{\prime })=\rho G_{V}L^{D}(t,t^{\prime })$.
Then, the equations for the mass and number become%
\begin{eqnarray}
\frac{dR\left( t,t^{\prime }\right) }{dt} &=&\frac{k_{g}G_{A}}{D\rho G_{v}}%
x,\;R(t,t)=R_{0} \\
\frac{d\mathcal{N}(t)}{dt} &=&Mk_{n}x+G_{A}k_{a}x\int_{0}^{t}R^{D}(t,t^{%
\prime })\frac{d\mathcal{N}(t^{\prime })}{dt^{\prime }}dt^{\prime },  \notag
\end{eqnarray}%
Note that the initial radius is related to the initial mass of a cluster by
where $m_{0}^{{}}=\rho G_{V}R_{0}^{D}$. Next, introduce the variable $u(t)=%
\frac{R(t,0)-R_{0}}{R_{0}}$. From 
\begin{equation*}
R(t,t^{\prime })=R_{0}+\frac{k_{g}G_{A}}{D\rho G_{v}}\int_{t^{\prime
}}^{t}x\left( t^{\prime \prime }\right) dt^{\prime \prime }
\end{equation*}%
we derive the following expression 
\begin{equation*}
R(t,t^{\prime })=R(t,0)-R(t^{\prime },0)+R_{0}=\left( u(t)-u(t^{\prime
})+1\right) R_{0}
\end{equation*}%
From 
\begin{equation*}
\frac{du}{dt}=\frac{1}{R_{0}}\frac{dR(t,0)}{dt}=\frac{k_{g}G_{A}}{D\rho G_{v}%
}x(t)\equiv \gamma x(t)
\end{equation*}%
and 
\begin{equation*}
\frac{dx}{dt}=\nu _{1}(n,x)-\frac{ds}{dt}=\nu _{1}(n,x)-\frac{ds}{du}\frac{du%
}{dt}
\end{equation*}%
we have already derived eqns~[\ref{model}] and [\ref{defgamma}]. However, we
still need to prove eqns~(\ref{su}-\ref{defab}).

Notice that the radius increases monotonically with time. We can therefore
replace time by $u(t)\ $and write $\mathcal{N}(t)=\frac{M}{m_{0}}\widetilde{%
\mathcal{N}}(u(t))$ for some function $\widetilde{\mathcal{N}}$, (the
prefactor being introduced to simplify later expressions). Using $\frac{d%
\mathcal{N}}{dt}=\frac{M}{m_{0}}\frac{d\widetilde{\mathcal{N}}}{du}\frac{du}{%
dt}$, the equation for the number of clusters can be rearranged to give%
\begin{equation}
\frac{d\widetilde{\mathcal{N}}(u)}{du}=\beta +\alpha \int_{0}^{u}\left(
u-u^{\prime }+1\right) ^{D-1}\frac{d\widetilde{\mathcal{N}}(u^{\prime })}{%
du^{\prime }}du^{\prime }  \label{NN}
\end{equation}%
with%
\begin{equation}
\alpha =\frac{DG_{V}k_{a}\rho R_{0}^{D}}{k_{g}}=\frac{Dk_{a}m_{0}}{k_{g}}%
,\quad \beta =\frac{m_{0}DG_{V}\rho R_{0}k_{n}}{G_{A}k_{g}}
\end{equation}%
An integration by parts gives 
\begin{equation}
\frac{d\widetilde{\mathcal{N}}(u)}{du}=\beta +\alpha \widetilde{\mathcal{N}}%
(u)+\left( D-1\right) \alpha \int_{0}^{u}\left( u-u^{\prime }+1\right) ^{D-2}%
\widetilde{\mathcal{N}}(u^{\prime })du^{\prime }.
\end{equation}%
where we assumed $\widetilde{\mathcal{N}}(0)=0$ as there is no solid present
at $t=0$. Continuing to differentiate gives%
\begin{align}
\frac{d^{2}\widetilde{\mathcal{N}}(u)}{du^{2}}& =\alpha \frac{d\widetilde{%
\mathcal{N}}(u)}{du}+\alpha \left( D-1\right) \widetilde{\mathcal{N}}(u)
\label{N1} \\
& +\alpha \frac{\left( D-1\right) !}{\left( D-3\right) !}\int_{0}^{u}\left(
u-u^{\prime }+1\right) ^{D-3}\widetilde{\mathcal{N}}(u^{\prime })du^{\prime }
\notag \\
\frac{d^{3}\widetilde{\mathcal{N}}(u)}{du^{3}}& =\alpha \frac{d^{2}%
\widetilde{\mathcal{N}}(u)}{du^{2}}+\alpha \left( D-1\right) \frac{d%
\widetilde{\mathcal{N}}(u)}{du}  \notag \\
& +\alpha \frac{\left( D-1\right) !}{\left( D-3\right) !}\widetilde{\mathcal{%
N}}(u)  \notag \\
& +\alpha \frac{\left( D-1\right) !}{\left( D-4\right) !}\int_{0}^{u}\left(
u-u^{\prime }+1\right) ^{D-4}\widetilde{\mathcal{N}}(u^{\prime })du^{\prime }
\notag
\end{align}%
where we used that 
\begin{equation}
\frac{\ud }{\ud x} \int_{a}^x f(x,y) \ud y=
f(x,x)+\int_{a}^x \frac{\ud f(x,y) }{\ud x}  \ud y
\end{equation}
for any function $f$ and constant $a$. The original integral equation is
therefore equivalent to a simple $D$-order differential,%
\begin{equation}
\frac{d^{D}\widetilde{\mathcal{N}}(u)}{du^{D}}=\alpha \sum_{j=1}^{D}\frac{%
\left( D-1\right) !}{\left( D-j\right) !}\frac{d^{D-j}\widetilde{\mathcal{N}}%
(u)}{du^{D-j}}
\end{equation}%
equation with boundary conditions%
\begin{eqnarray}
\widetilde{\mathcal{N}}(0) &=&0,\quad \left. \frac{d\widetilde{\mathcal{N}}%
(u)}{du}\right| _{0}=\beta   \label{bc} \\
\left. \frac{d^{j}\widetilde{\mathcal{N}}(u)}{du^{j}}\right| _{0} &=&\alpha
\sum_{i=1}^{j-1}\frac{\left( D-1\right) !}{\left( D-i\right) !}\left. \frac{%
d^{j-i}\widetilde{\mathcal{N}}(u)}{du^{j-i}}\right| _{0},\;j=2...D-1  \notag
\end{eqnarray}%
The solution is a sum of exponentials, 
\begin{equation}
\widetilde{\mathcal{N}}(u)=\sum_{j=1}^{D}a_{j}\exp \left( \lambda
_{j}u\right) ,  \label{sumexp}
\end{equation}%
where the constants $\lambda _{i}$ are the roots of 
\begin{equation}
\lambda ^{D}-\alpha \sum_{j=1}^{D}\frac{\left( D-1\right) !}{\left(
D-j\right) !}\lambda ^{D-j}=0
\end{equation}%
Using Eq.(\ref{bc}),  the coefficients satisfy%
\begin{eqnarray}
0 &=&\sum_{j=1}^{D}a_{j} \\
\beta  &=&\sum_{j=1}^{D}\lambda _{j}a_{j},\;\text{if }D>1  \notag \\
\alpha \beta  &=&\sum_{j=1}^{D}\lambda _{j}^{2}a_{j},\;\text{if }D>2  \notag
\end{eqnarray}%
and so on. In general, solution for the coefficients requires solving this
system. For example, for the most interesting cases of $D=2$ and $D=3$ , the
solution  is%
\begin{eqnarray}
a_{i} &=&\beta \frac{1}{2\lambda _{i}-\lambda _{1}-\lambda _{2}},\;D=2 \\
a_{i} &=&\beta \frac{\alpha -\sum_{j\neq i}\lambda _{j}}{\prod\limits_{j\neq
i}\left( \lambda _{i}-\lambda _{j}\right) },\;D=3  \notag
\end{eqnarray}%
Finally, the mass fraction of solid is%
\begin{eqnarray}
s(t) &=&M^{-1}\int_{0}^{t}m(t,t^{\prime })\frac{d\mathcal{N}(t^{\prime })}{%
dt^{\prime }}dt^{\prime } \\
&=&\frac{G_{V}\rho }{m_{0}}R_{0}^{3}\int_{0}^{u(t)}\left( u(t)-u^{\prime
}+1\right) ^{D}\frac{d\widetilde{\mathcal{N}}(u^{\prime })}{du^{\prime }}%
du^{\prime }  \notag
\end{eqnarray}%
or $s(t)=\widetilde{s}\left( u(t)\right) $ with%
\begin{equation}
\widetilde{s}\left( u\right) =\int_{0}^{u}\left( u-u^{\prime }+1\right) ^{D}%
\frac{d\widetilde{\mathcal{N}}(u^{\prime })}{du^{\prime }}du^{\prime }
\end{equation}%
as $G_{V}\rho R_{0}^{3}/m_{0}=1$. Differentiating gives%
\begin{equation}
\frac{d}{du}\widetilde{s}\left( u\right) =\frac{d\widetilde{\mathcal{N}}(u)}{%
du}+D\int_{0}^{u}\left( u-u^{\prime }+1\right) ^{D-1}\frac{d\widetilde{%
\mathcal{N}}(u^{\prime })}{du^{\prime }}du^{\prime }
\end{equation}%
Substituting from eqn~(\ref{NN}),%
\begin{equation}
\frac{d}{du}\widetilde{s}\left( u\right) =\frac{d\widetilde{\mathcal{N}}(u)}{%
du}+\frac{D}{\alpha }\left( \frac{d\widetilde{\mathcal{N}}(u)}{du}-\beta
\right) 
\end{equation}%
giving%
\begin{equation}
\widetilde{s}\left( u\right) =\left( 1+\frac{D}{\alpha }\right) \widetilde{%
\mathcal{N}}(u)-D\frac{\beta }{\alpha }u.  \label{sunu}
\end{equation}%
Insertion of eqn~(\ref{sumexp}) into eqn~(\ref{sunu}) results in eqn~(\ref%
{su}).

\subsection{Derivation of eqn~(\ref{model-rate})}

In order to make contact with the phenomenological model, it is useful to
develop the solution for the mass fraction as a power series. From eqn~(\ref%
{N1}) and the boundary conditions, it is easily shown that%
\begin{equation}
\widetilde{\mathcal{N}}(u)=\sum_{n=0}^{\infty }c_{n}u^{n}
\end{equation}%
with%
\begin{align}
c_{0}& =0,\quad c_{1}=\beta ,\quad c_{2}=\frac{1}{2}\alpha \beta ,\quad 
\text{and for }n\geq 3 \\
c_{n}& =\frac{\alpha }{n}c_{n-1}+\frac{2\alpha }{n\left( n-1\right) }c_{n-2}+%
\frac{2\alpha }{n\left( n-1\right) \left( n-2\right) }c_{n-3}.  \notag
\end{align}%
Thus, 
\begin{eqnarray}
\widetilde{s} &=&\left( 1+\frac{3}{\alpha }\right) \sum_{n=1}^{\infty
}c_{n}u^{n}-3\frac{\beta }{\alpha }u \\
&=&\beta u+\frac{1}{2}\beta \left( \alpha +3\right) u^{2}+\frac{1}{6}\beta
\left( \alpha +3\right) \left( \alpha +2\right) u^{3}+...  \notag
\end{eqnarray}%
We can invert this power series using Lagrange inversion theorem which
states that 
\begin{equation}
u(s)=\sum_{n=1}^{\infty }\frac{{}^{n-1}}{{}u^{n-1}}\left. \left( \frac{u}{%
s(u)}\right) ^{n}\right| _{u=0}\frac{s^{n}}{n!}
\end{equation}%
and results in 
\begin{equation*}
u=\frac{1}{\beta }\widetilde{s}-\frac{\left( \alpha +3\right) }{2\beta ^{2}}%
\widetilde{s}^{2}+\frac{\left( 2\alpha +7\right) \left( \alpha +3\right) }{%
6\beta ^{3}}\widetilde{s}^{3}+...
\end{equation*}%
Finally, to fully eliminate $u$ in favor of $s$ we need%
\begin{align}
\frac{d}{du}\widetilde{s}\left( u\right) & =\beta +\beta \left( \alpha
+3\right) u+\frac{1}{2}\beta \left( \alpha +3\right) \left( \alpha +2\right)
u^{2}+\ldots   \notag \\
& =\beta +\left( \alpha +3\right) \widetilde{s}-\frac{1}{2}\frac{(\alpha +3)%
}{\beta }\widetilde{s}^{2}+...  \notag \\
&
\end{align}%
Substitution of this relation into the model, eqn~(\ref{model}), gives the
rate-like form, eqn~(\ref{model-rate}).

\subsection{Size distribution}

Let $\mathcal{N}(R,t)$ be the number of clusters of radius $R$ or smaller.
Since the clusters grow monotonically, a cluster of radius $R$ at time $t$
was nucleated at some definite time $t^{\prime }(R)<t$. Hence, the total
number of clusters with radius less than $R$ is the total number of clusters
minus the number with radius greater than $R$ which is to say the total
number minus the number already present at time $t^{\prime }(R)$, 
\begin{equation}
\mathcal{N}\left( R,t\right) =\left[ \mathcal{N}\left( t\right) -\mathcal{N}%
\left( t^{\prime }\left( R\right) \right) \Theta \left( R_{m}\left( t\right)
-R\right) \right] \Theta \left( R-R_{0}\right)
\end{equation}%
where the step function $\Theta \left( R-R_{0}\right) $ enforces the
condition that there are no clusters smaller than $R_{0}$ and the step
function $\Theta \left( R_{m}\left( t\right) -R\right) $ is required since
there is a maximal size corresponding to a cluster nucleated at time $t=0$
(assuming there are no clusters present at $t<0$). Hence, when $R>R_m$, $%
\mathcal{N}(R,t)$ is simple equal to the total number of clusters $\mathcal{N%
}(t)$. Now, the problem is to find $t^{\prime }(R)$. It is sufficient to
note that $R=R(t,t^{\prime })=\left( u(t)-u(t^{\prime })+1\right) R_{0}$.
Since $\mathcal{N}\left( t^{\prime }\left( R\right) \right) = \frac{M}{m_0} 
\widetilde{\mathcal{N}}\left( u(t^{\prime }\left( R\right) )\right) = \frac{M%
}{m_0} \widetilde{\mathcal{N}}\left( u+1-\frac{R}{R_{0}}\right) $. It then
follows that%
\begin{align}
&\mathcal{N}\left( R,t\right) = \Theta \left( R-R_{0}\right) \frac{M}{m_0}
\times \\
&\left( \widetilde{\mathcal{N}}\left( u(t)\right) -\widetilde{\mathcal{N}}%
\left( u\left( t\right) +1-\frac{R}{R_{0}}\right) \Theta \left( R_{m}\left(
t\right) -R\right) \right)  \notag
\end{align}%
Finally, note that $R_{m}$ is found by taking $t^{\prime }=0$ giving $%
R_{m}=R(t,0)=\left( u(t)+1\right) R_{0}$ so that%
\begin{align}
&\mathcal{N}\left( R,t\right) =\Theta \left(R- R_{0} \right) \frac{M}{m_0}
\times \\
&\left( \widetilde{\mathcal{N}}\left( u(t)\right) -\widetilde{\mathcal{N}}%
\left( u\left( t\right) +1-\frac{R}{R_{0}}\right) \Theta \left( \left(
u(t)+1\right) R_{0}-R\right) \right)  \notag
\end{align}%
This expression simply means that the number of clusters of size $R$ or
smaller is the total number of clusters created since the time $t^{\prime
}(R)$ at which clusters of size $R$ were created.

\subsection{No cooperativity}

The limit of no cooperativity, $k_{a}=\alpha=0 $, is not easy to extract
from the general solution. A simpler approach is to return to eqn~(\ref{NN})
which, in this limit, becomes 
\begin{equation}
\frac{d\mathcal{N}(t)}{dt}=Mk_{n} x = M k_n \gamma^{-1} \frac{du}{dt}= \frac{
M\rho R_{0} k_n}{k_{g}}\frac{du}{dt}
\end{equation}%
or%
\begin{equation*}
\frac{d\frac{m_{0}}{M}\mathcal{N}(t)}{dt}= \frac{d \widetilde{\mathcal{N}}(t)%
}{dt}= \frac{ m_0 \rho R_{0} k_n}{k_{g}}\frac{du}{dt}= \beta \frac{du}{dt}
\end{equation*}%
so that $\widetilde{\mathcal{N}}(u)=\beta u$. Substituting this into the
result for the size distribution gives%
\begin{align}
&\mathcal{N}\left( R,t\right) =\beta \left( \frac{M}{m_{0}}\right) \Theta
\left(R- R_{0}\right) \times \\
&\left( u(t)-\left( u\left( t\right) +1-\frac{R}{R_{0}}\right) \Theta \left(
\left( u(t)+1\right) R_{0}-R\right) \right)  \notag
\end{align}%
The total mass fraction can now be calculated since the mass of a cluster of
radius $R$ is simply $G_{V}R^{D}\rho $ and the number of clusters with
radius between $R$ and $R+dR$ is $\frac{dn(R,t)}{dR}dR $ and $\frac{dn(R,t)}{%
dR}=\beta \frac{M}{m_0}\frac{1}{R_0} \Theta \left(R- R_{0}\right) \Theta
\left( \left( u(t)+1\right) R_{0}-R\right)$. Summing over all radii gives 
\begin{eqnarray}
s\left( t\right) &=&\frac{1}{M}\int_{0}^{\infty }G_{V}R^{D}\rho \frac{d%
\mathcal{N}\left( R,t\right) }{dR}dR \\
&=&\beta \frac{1}{M}\left( \frac{M}{m_{0}}\right) \int_{R_{0}}^{\left(
u(t)+1\right) R_{0}}\frac{1}{R_{0}}G_{V}R^{D}\rho dR  \notag \\
&=& \frac{\beta }{D+1}\left( \left( u(t)+1\right) ^{D+1}-1\right)  \notag
\end{eqnarray}%
In this limit, our result thus reduces to a form that occurs in the theory
of crystallization\cite{Mullin}. \bigskip

\bibliographystyle{apsrev}

\end{document}